\documentclass[prd,showpacs,showkeys,preprint,12pt]{revtex4}
\usepackage{amsmath,amsthm,amsfonts,amssymb}
\usepackage[mathcal]{eucal}
\usepackage{mathrsfs}
\usepackage{graphicx}
\usepackage{dcolumn}
\usepackage{latexsym}
\usepackage{epsfig}
\usepackage{bm}
\usepackage[all]{xy}
\newcommand{\ie}{\begin{equation}}
\newcommand{\fe}{\end{equation}}

\newcommand\fverb{\setbox\fverbbox=\hbox\bgroup\verb}
\newcommand\fverbdo{\egroup\medskip\noindent%
            \fbox{\unhbox\fverbbox}\ }
\newcommand\fverbit{\egroup\item[\fbox{\unhbox\fverbbox}]}
\newbox\fverbbox

\def\text#1{\mbox{#1}}

\begin{document}

\title{A 6D standing wave Braneworld.}
\author{L. J. S. Sousa$\;^{a,b}$}
\author{J. E. G. Silva$\;^{a}$}
\author{C. A. S. Almeida$\;^{a}$}
\address{$\;^{a}$Departamento de F\'{\i}sica, Universidade Federal do
Cear\'a \\ Caixa Postal 6030, CEP 60455-760, Fortaleza, Cear\'a, Brazil}
\address{$\;^{b}$Instituto Federal de Educa\c{c}\~{a}o, Ci\^{e}ncia e Tecnologia do Cear\'a (IFCE)\\ Campus Canind\'{e} - Canind\'{e}, Cear\'a, Brazil}

\begin{abstract}

We constructed a six-dimensional version of the standing wave model with an anisotropic 4-brane generated by a phantom-like scalar field.  The model represents a braneworld where the compact (on-brane) dimension is
assumed to be sufficiently small in order to describe our universe (hybrid compactification). The proposed geometry of the brane and its transverse manifold is non-static, unlike the majority of braneworld models
presented in the literature. Furthermore, we have shown that the zero-mode scalar field is localized around the brane. While in the string-like defect the scalar field is localized on a brane with decreasing warp factor,
here it was possible to perform the localization with an increasing warp factor.

\end{abstract}


\maketitle

\section{Introduction}

The idea that our world is a brane embedded in a higher-dimensional space-time has attracted the attention of the physics community in the last years, basically because the braneworld idea has
brought solution for some intriguing problems in the Standard Model (SM), as the hierarchy problem. The mainly theories that carrier the braneworld basic idea are the one first proposed by
Arkani-Hamed, Dimopoulos and Dvali \cite{I.Antoniadis1998, N.Arkani-Hamed1999, N.Arkani-Hamed1999a} and the so-called, Randall-Sundrum (RS) model \cite{Randall1999a,Randall1999}.

One important feature of these models is the
assumption that all matter fields are constrained to propagate only on the brane, whereas gravity is free to propagate in the extra dimensions. However the presupposition that
the Standard Model fields are initially trapped on the brane is not so obvious in this framework. In this way, it is interesting to seek alternative field theoretic localization mechanisms
in braneworld scenarios \cite{Oda:2000zc,Oda2000a}. Therefore, the property of a model to localize fields has been used as a parameter to evaluate it as a potential candidate of our universe.

Particularly the RS model, which is a 5D theory, was quickly generalized to six dimensions \cite{Gregory:1999gv,Chen:2000at,Cohen:1999ia,Olasagasti:2000gx,Gherghetta:2000qi,Ponton:2000gi,
Oda:2000zc,Oda2000a,Giovannini:2001hh,Tinyakov:2001jt,Kanno:2004nr,Vinet:2004bk,Cline:2003ak,Papantonopoulos:2007fk,
Navarro:2003vw,Navarro:2004di,Papantonopoulos:2005ma,deCarlos:2003nq,Gogberashvili:2001jm,Koley2007,Torrealba,Silva:2011yk,Luis2012}. On the other hand, the scenarios where the brane has cylindrical symmetry are the so-called string-like braneworlds. There is a broad variety of these models as the global string \cite{Oda:2000zc,Gregory:1999gv,Olasagasti:2000gx},
the local string \cite{Gherghetta:2000qi,Ponton:2000gi,Giovannini:2001hh}, the thick string \cite{Navarro:2003vw,Navarro:2004di,Papantonopoulos:2005ma,Papantonopoulos:2007fk,Vinet:2004bk,Kanno:2004nr} and the
supersymmetric cigar-universe \cite{deCarlos:2003nq}.

There exist a vast literature concerned to the localization of fields both in 5D \cite{Kehagias2001,Cruz2009,W.T.Cruz2011,Tahim2009,Merab2011a,Merab2011b} and in 6D braneworld
\cite{Gherghetta:2000qi,Oda:2000zc,Oda2000a,Giovannini:2001hh,Torrealba,Silva:2011yk,Luis2012}. However, to the best of our knowledge, there is not yet a purely analytical geometry that traps all the SM fields through only the gravitational field interaction.
Hence, the quest for a model which is both analytical and that localizes all the SM fields by means of only the gravitational interaction, is in our point of view, a reasonable justification to continue studying field localization in different braneworld models.

In this spirit, it has been proposed some braneworld scenarios with non-standard transverse manifold. Randjbar-Daemi and
Shaposhnikov has assumed it as a Ricci-flat or an homogeneous space and they obtained trapped massless gravitational modes and chiral fermions as well \cite{RandjbarDaemi:2000ft}. Kehagias proposed a conical tear-drop whose conical singularity drains the vacuum energy explaining the small value of the cosmological constant \cite{Kehagias:2004fb}. Gogberashvili \textit{et al.} have achieved three-generation for fermions on a $3$-brane whose transverse space has an apple shape \cite{Gogberashvili:2007gg}. Other examples are the torus \cite{Duan:2006es}, a football-shape space \cite{Garriga:2004tq} and  smoothed versions of the conifold, the resolved \cite{VazquezPoritz:2001zt} and deformed one\cite{Brummer:2005sh, Firouzjahi:2005qs, Noguchi:2005ws}.

In this paper we considered a bulk phantom-like scalar field in a 6D braneworld as a source of the brane. The phantom is a scalar field whose sign of the kinetic term in its Lagrangian is negative. This model exhibits
instability issues, as unbounded negative energy, but phenomenologically it has been useful in different scenarios \cite{Koley2007}. In cosmology, the phantom scalar is a candidate to explain dark energy and the accelerated expansion of the universe \cite{Caldwell et al}.

The use of bulk scalar fields to generate branes was introduced by Goldemberg and Wise
\cite{wd.goldberger-prl83, wd.goldberger-prd60}, and it has been largely studied in the literature \cite{o.dewolfe-prd62, rn.mohapatra-prd62, p.kanti-plb481, jm.cline-prd64, jm.cline-plb495, a.flachi-npb610}.
The phantom field has been implemented in the six-dimensional context by Koley and Kar \cite{Koley2007}, where it has shown the localization of gauge fields.

In the present model the field is
not a phantom in the sense described above. Despite it has the "wrong-sign" characteristic of the phantom its energy is not ever negative and  then, there are no energy problems. Thus, it is more convenient call this source as a "phantom-like scalar".

This exotic source was first proposed in five dimensions, the so-called standing-wave braneworld. This is a completely anisotropic one brane model whose phantom-scalar is equivalent to a Weyl scalar \cite{Merab2011a,Merab2011b,Gogberashvili:2010yf}. The Weyl scalar, in its turn, is used in an extension of the RS model, the so-called pure-gravity braneworld. In the standing-wave approach it was possible to
localize various fields, albeit it was not possible trap right-handed fermions \cite{Merab2011a,Merab2011b}. Besides, the study of massive modes was not addressed in this model.

The model built in this work consists of a 6D braneworld with an anisotropic 4-brane, where the compact (on-brane) dimension is assumed to be sufficiently small in order to accomplish our universe (hybrid compactification). Anisotropic six-dimensional braneworlds have already been addressed before in the thick string-like models \cite{Navarro:2003vw,Navarro:2004di,Papantonopoulos:2005ma,Papantonopoulos:2007fk,Kanno:2004nr}, where there are different warp factors for different directions. Nonetheless, it is assumed the brane has a cosmological constant which turned it a locally isotropic manifold. However, the standing-wave braneworld is a completely anisotropic manifold, unless for some points called the $AdS$ islands \cite{Merab2011a,Merab2011b}.

Another appreciable feature of our model is its dynamics, in the sense that both metric and the bulk phantom-like scalar field are time dependent. Indeed, as in RS model, the most string-like models are static \cite{Cline:2003ak,Cohen:1999ia,deCarlos:2003nq,Gherghetta:2000qi,Navarro:2003vw,Navarro:2004di,Oda:2000zc,Papantonopoulos:2005ma,
Papantonopoulos:2007fk,Ponton:2000gi,Silva:2011yk,Olasagasti:2000gx,Tinyakov:2001jt,Vinet:2004bk}.
On the other hand, Gregory has proposed a time-dependent string-like brane leading to gravity localization \cite{Gregory:1999gv}. The difference here is that both the $3$-brane and the transverse manifold are time-dependent. This allows the exterior geometry of the brane to reflect the changes of the brane, a key property of the string-like defects \cite{Chen:2000at,Olasagasti:2000gx,Ponton:2000gi,Giovannini:2001hh}.

In spite of the complexity due the time-dependence and anisotropy, we have obtained analytical solution for the warp factor and the phantom field.
The bulk is everywhere smooth converging asymptotically to an $AdS_{6}$ manifold.
From these solutions we have analyzed some properties of the source. The energy-momentum components have compact support around the origin what means that this geometry is created by a local source. Nevertheless, they do not satisfy the energy conditions, which turns this an exotic scenario, as expected. However, even the thin string-like braneworlds do not obey the dominant energy-condition \cite{Gherghetta:2000qi,Tinyakov:2001jt}.  Besides, in the phantom 6D braneworld the energy momentum tensor violates all of
energy conditions \cite{Koley2007}. Although the bulk
spacetime obtained in that setup is not dynamical stable, the authors argued that it also occurs
in other models \cite{Koley2007}.

Once studied the geometry in details we have concerned with the behavior of a minimally coupled scalar field in this model. As in the string-like model and the 5D standing-wave this field has its massless mode trapped to the brane. The difference between the string-like and the 6D standing-wave approach is that in the former the scalar field is trapped for a decreasing warp factor whereas in the latter there is a localized mode for an increasing warp factor.

This work is organized as follows: in section $2$ we described the model and solved the Einstein and scalar field equations obtaining the general static expression for the phantom-like scalar.
In section 3 we have found the non-static standing waves and we have discussed its mainly characteristics. The localization of the zero mode scalar field have been done in the section $4$. In the section $5$, some final remarks and conclusions were outlined.

\section{The model}

We begin by the general action in 6-dimensional space-time composed by the standard Einstein-Hilbert action and a bulk massless scalar field action which is time dependent and minimally coupled to gravity

\begin{equation}\label{action}
S=\frac{1}{2\kappa_{6}^{2}} \int d^{6}x \sqrt{-\;^{(6)}g}\Big[(R - 2\Lambda) + g^{AB}\nabla_{A} \Phi \nabla_{B}\Phi - V(\Phi)\Big],
\end{equation}
where $\kappa _{6}$ is the 6-dimensional gravitational constant, and $\Lambda$ is the bulk cosmological constant. The "wrong" signal in front of the kinetic term of the scalar field action characterizes it as a phantom-like field.

The variation of this action with respect to the metric and the field $\Phi$ give us the following  equations of motion
\begin{equation} \label{Einstein}
R_{MN}-\frac{1}{2} g_{MN}R = -\Lambda g_{MN} + \kappa_{6}^{2}T_{MN},
\end{equation}

\begin{equation} \label{klein-gordon1}
\frac{1}{\sqrt{-g}}\partial_{M} \Big\{ \sqrt{-g}\;g^{MN} \partial_{N}\Phi \Big\} = \frac{\partial V}{\partial \Phi},
\end{equation}
where $M,N,...$ denote D-dimensional space-time indices and $T_{MN}$ is the energy-momentum tensor.

We will begin by considering the general metric ansatz
\begin{equation}
\label{gen-metric}
ds^{2}= e^{A}\left( dt^{2} - e^{u}dx^{2} - e^{u}dy^{2} - e^{-3u}dz^{2} \right) - dr^{2} - R_{0} ^{2}e^{B + u}d\theta^{2},
\end{equation}
where  $A(r)$ and $B(r)$ are functions of $r$, only, and $u$ is function of $r$ and $t$. This ansatz generalizes the global string like defect considered in Refs. \citep{Oda:2000zc, Gregory:1999gv}, for instance.

For a scalar field the energy-momentum tensor may be written as
\begin{equation} \label{gen-EM}
T_{MN} = \partial_{M}\Phi\partial_{N}\Phi - g_{MN}\left(\frac{1}{2} \partial^{C}\Phi\partial_{C}\Phi + V(\Phi) \right),
\end{equation}
from which the Einstein equation may be rewritten as
\begin{equation}
\label{Einstein3}
R_{MN} = \kappa_{6} ^{2}\partial_{M}\Phi\partial_{N}\Phi + \frac{1}{2} g_{AB} (\kappa_{6} ^{2} V(\Phi)  + \Lambda_{6}).
\end{equation}

From Eqs. $(\ref{Einstein})$ and $(\ref{Einstein3})$ the non zero components of the Ricci tensor, for $V(\Phi)=0$, are given as
\begin{eqnarray}
\label{Ricci-xx1}
\lefteqn{R_{xx} = R_{yy} = \left(-\frac{1}{4} e^{A + u}\right)}\nonumber\\
& &\left( 4A^{'2} + 2A^{''} +A^{'}B^{'} -2 e^{-A}\ddot{u} + (4A^{'} + B^{'})u^{'} +2 u^{''}\right) \nonumber\\ & & = -\frac{1}{2} e^{A + u} \Lambda_{6},
\end{eqnarray}
\begin{eqnarray}
\label{Ricci-zz1}
\lefteqn{ R_{zz} = \left(\frac{1}{4} e^{A -3 u}\right)}\nonumber\\
& &\left( -4A^{'2} - 2A^{''} -A^{'}B^{'} -6 e^{-A}\ddot{u} + 3(4A^{'} + B^{'})u^{'} +6 u^{''}\right) \nonumber\\ & & = -\frac{1}{2} e^{A -3 u} \Lambda_{6},
\end{eqnarray}
\begin{equation}
 \label{Ricci-tt1}
R_{tt} = \frac{1}{4} e^{A}\left( 4A^{'2} + 2A^{''} +A^{'}B^{'} -12e^{-A}\dot{u}^2\right)= \kappa_{6} ^{2} \dot{\Phi}^{2} +\frac{1}{2} e^{A } \Lambda_{6},
\end{equation}
\begin{equation}
 \label{Ricci-rt1}
R_{rt} = \frac{1}{4}\dot{u}(A^{'} - B^{'} - 12 u^{'})= \kappa_{6} ^{2}\dot{\Phi} \Phi^{'},
\end{equation}
\begin{eqnarray}
\label{Ricci-rr1}
\lefteqn{ R_{rr} = \left(\frac{1}{4} e^{A -3 u}\right)}\nonumber\\
& &\left( -4A^{'2} -B^{'2} - 8A^{''} - 2B^{''} +2(A^{'} - B^{'})u^{'} -12 u^{'2}\right) \nonumber\\ & & =\kappa_{6} ^{2} \Phi^{'2} -\frac{1}{2}  \Lambda_{6},
\end{eqnarray}
\begin{eqnarray}
\label{Ricci-teta1}
\lefteqn{ R_{\theta \theta} = \left(-\frac{1}{4}R_{0} ^{2} e^{B + u}\right)}\nonumber\\
& &\left( B^{'2} + 2B^{''} + 4 A^{'}B^{'} -2 e^{-A}\ddot{u} + (4A^{'} + B^{'})u^{'} +2 u^{''}\right) \nonumber\\ & & = -\frac{1}{2} R_{0} ^{2} e^{B + u} \Lambda_{6},
\end{eqnarray}
which may be simplified to
\begin{equation} \label{Einstein-simplif1}
- e^{-A}\ddot{u} + \frac{1}{2}(4A^{'} + B^{'})u^{'} + u^{''}=0,
\end{equation}
\begin{equation}
 \label{Ricci-rt1}
R_{rt} = \frac{1}{4}\dot{u}(A^{'} - B^{'} - 12 u^{'})= \kappa_{6} ^{2}\dot{\Phi} \Phi^{'},
\end{equation}
\textbf{and}
\begin{eqnarray}
\label{Einstein-simplif2}
\lefteqn{\frac{1}{4} \left( - 6A^{''} +5 A^{'}B^{''} + \right)}\nonumber\\
& & + \frac{1}{4} \left((6A^{'} - B^{'})u^{'} + 2u^{''} - 2 e^{-A}\ddot{u} -12(e^{-A}\dot{u}^2 + u^{'2}) \right) \nonumber\\
& &  =\kappa_{6} ^{2} (e^{-A}\dot{\Phi}^{2}+ \Phi^{'2}) +\frac{1}{2}  \Lambda_{6}.
\end{eqnarray}
For $A = B$, Eq. $(\ref{Ricci-rt1}$) implies that $\Phi$ has to be a phantom-like scalar (minus sign in front of the kinetic term in Lagrangian) which relates to the function $u$ as $\Phi \equiv \sqrt{\frac{3}{\kappa_{6} ^{2}}}u$. In this case the equations for the function $u$ and the warp factor $A$ are given, respectively, as
\begin{equation} \label{Einstein-simplif3}
- e^{-A}\ddot{u} + \frac{5}{2}A^{'} u^{'} + u^{''}=0,
\end{equation}

\begin{equation} \label{warp-factor1}
A^{''} (r) - \frac{5}{6} A^{'2} + \frac{\Lambda_{6}}{3} = 0.
\end{equation}
The solution to the last equation is than, unless integration constants,
\begin{equation} \label{warp-factor1-sol}
A (r)=  - \frac{6}{5}\log \left[\cosh \left( \sqrt{\frac{5 \Lambda_{6}}{18}}r \right)\right].
\end{equation}
This warp factor represents a thick brane and the model proposed here generalizes the 5D standing wave braneworld not only because we are working in six dimensions but in the sense that here we have a thick brane instead of a thin brane.

It is interesting to note that the field $\Phi$, in general, is do not required to be a phantom one. This solution is a special case when we consider $A=B$, as we saw above. For $A \neq B$, in general, it is difficult to solve the system of equations, Eq. (\ref{Ricci-xx1}) to Eq. (\ref{Ricci-teta1}). However one could find the function $u$, the warp factors $A$ and $B$ and the components $\dot{\Phi}^2, \Phi^{'2}$ of the bulk scalar and analyzes the energy conditions. This job will be leave for another work because here we are interested to study field localization in this scenario.

Hereinafter, we shall consider $A(r) B(r)= 2 a r$ and use the metric ansatz
\begin{equation}
\label{metric}
ds^{2}= e^{2 a r}\left( dt^{2} - e^{u}dx^{2} - e^{u}dy^{2} - e^{-3u}dz^{2} \right) - dr^{2} - R_{0} ^{2}e^{2 a r + u}d\theta^{2},
\end{equation}
where $a\in R$ with dimension $[a]=M$, $0\leq r < \infty$, $0\leq \theta < 2\pi$ and $u = u(r,t)$ is function of $t$ and $r$, only. The compact dimension $\theta$ lives on the brane, i.e., $\theta$ is a brane coordinate for $r=0$.

The metric ansatz $(\ref{metric})$ is a combination of the 6D warped braneworld model through the $e^{2ar}$ term \cite{Gregory:1999gv,Oda:2000zc,Oda2000a,Koley2007} plus an anisotropic (t, r)-dependent warping of the brane coordinates, x, y, z, through the terms $e^{u(t,r)}$ and $e^{ -3 u(t,r)}$, which means a six dimensional generalization of the 5D standing-wave braneworld model \cite{GOGBERASHVILI1,Gogberashvili:2010yf,Merab2011a,Merab2011b,Merab2012}.

The metric $(\ref{gen-metric})$ and, naturally, the metric $(\ref{metric})$, are examples of the general metric
\begin{equation}
\label{generalmetric}
ds^{2}=g_{\mu\nu}(x,r)dx^{\mu}dx^{\nu} - dr^{2} - L(x,r)d\theta^{2},
\end{equation}
proposed by many authors in the thick brane model relevants for cosmological reasons \cite{Kanno:2004nr,Navarro:2003vw,Navarro:2004di,Papantonopoulos:2005ma,Papantonopoulos:2007fk,Vinet:2004bk}.

It is worthwhile to mention that the metric $(\ref{metric})$ has an time-dependent angular component. This allows some geometrical properties of the transverse manifold, as its deficit angle, to evolve in time. Hence, the evolution of the brane alters the geometry outside the brane.

Another noteworthy property of this scenario is that, for $u=0$, the metric $(\ref{metric})$ is the same of the thin string-like defects \cite{Chen:2000at,Olasagasti:2000gx,Gherghetta:2000qi,Ponton:2000gi,Oda:2000zc,Oda2000a}. Since these points form a discrete set, in these points the geometry is the called an $AdS$ island.

From the above mentioned properties, the function $u(r,t)$ can be regarded as a correction of the string-like models leading to an anisotropic, time-dependent and thick braneworld that we shall call $6$D standing-wave.

From metric  $(\ref{metric})$ and choosing $V(\Phi)= 0$ the scalar field equation $(\ref{klein-gordon1})$ reads
\begin{equation}
\label{klein-gordon}
\frac{1}{\sqrt{-g}}\partial_{M}\left(\sqrt{-g}g^{MN}\partial_{N} \Phi \right) = e^{-2\textit{a}r}\ddot{\Phi} - 5\textit{a} \Phi^{'} - \Phi^{''} = 0,
\end{equation}
where $g$ is the bulk metric determinant. The scalar phantom field energy-momentum tensor is given by
\begin{equation} \label{E-M}
T_{MN} = -\partial_{M}\Phi\partial_{N}\Phi + \frac{1}{2} g_{MN}\partial^{C}\Phi\partial_{C}\Phi.
\end{equation}
For further reference it is important to write here the bulk curvature scalar
\begin{equation} \label{curvature}
R = 30 a^{2} - 3 e^{-2 a r} \dot{u}^{2} + u^{'2}.
\end{equation}

From Eq. $(\ref{Einstein})$ the non-zero components of the Ricci tensor are
\begin{equation}
\label{Ricci-xx}
R_{xx} = \frac{1}{2} e^{2\textit{a}r + u}\left( -10\textit{a}^2 + e^{-2\textit{a}r}\ddot{u} -5\textit{a}u^{'} - u^{''}\right) = R_{yy},
\end{equation}
\begin{equation}
\label{Ricci-zz}
R_{zz} = \frac{3}{2} e^{2\textit{a}r -3u}\left( -\frac{10}{3}\textit{a}^2 - e^{-2\textit{a}r}\ddot{u} + 5\textit{a}u^{'} + u^{''}\right),
\end{equation}
\begin{equation}
 \label{Ricci-tt}
R_{tt} = \frac{1}{4} \left( 20\textit{a}^2 e^{2\textit{a}r} -12\dot{u}^2\right),
\end{equation}
\begin{equation}
 \label{Ricci-rt}
R_{rt} = -3\dot{u}u^{'},
\end{equation}
\begin{equation}
 \label{Ricci-rr}
R_{rr} = \frac{1}{4}\left( -20\textit{a}^{2} - 12u^{'2}\right),
\end{equation}
\begin{equation}
 \label{Ricci-teta}
R_{\theta\theta} = \frac{R_{0} ^{2}}{4}e^{2 a r + u}\left( -20\textit{a}^{2} - 10a u^{'} - 2u^{''} + 2 e^{-2ar} \ddot{u} \right).
\end{equation}

From the energy-momentum tensor $(\ref{E-M})$ we may rewrite the Einstein equation $(\ref{Einstein3})$ in the following simpler form
\begin{equation}
\label{Einstein2}
R_{MN} = -\partial_{M}\sigma\partial_{N}\sigma + \frac{1}{2} g_{AB}\Lambda_{6}
\end{equation}
where $\sigma = \sqrt{\kappa}\Phi$.

Comparing Eq. $(\ref{Ricci-rt})$ and the Ricci components coming from Eq. $(\ref{Einstein2})$ we need to impose that $\sigma = \sqrt{3}u$ to obtain a solution. This imply that the $u$ field, according to $\sigma = \sqrt{\kappa}\Phi$, needs to satisfies the relation $(\ref{klein-gordon})$. This is possible if, by comparing Eq. $(\ref{Ricci-xx})$ with $R_{xx}$ in Eq. $(\ref{Einstein2})$, we do
\begin{equation}
\Lambda_{6} = 10\textit{a}^{2}.
\end{equation}

Therefore, the differential equation for the $u$, which is a simplification of Eq. $(\ref{Einstein-simplif3})$, is given by
\begin{equation}
 \label{u-equation}
 e^{-2 a r}\ddot{u}(r,t) - 5 a u^{'}(r,t) - u^{''}(r,t) = 0,
\end{equation}
where prime and dots mean differentiation with respect to $r$ and $t$, respectively. In order to solve equation $(\ref{u-equation})$ we proceed by separating the variable as follows
\begin{equation}
\label{u-variable-sepa1}
u(r,t) = g(t) \rho(r),
\end{equation}
where we require that $g(t)$ satisfies the following basic requisites: $\dot{g} \propto g$; $\ddot{g} \propto g$. In other words, $g(t)$ may be an function of time in the form of an exponential, or sine, or cosine, or hyperbolic sine, or hyperbolic cosine. Then, assuming that $g(t)$ satisfies some of these requirements, the equation to the new variable $\rho(r)$ reads
\begin{equation}
 \label{ro-equation}
\rho^{''}(r) + 5 a \rho^{'}(r) + \alpha^{2}e^{-2 a r}\rho(r) = 0,
\end{equation}
where $\alpha$ is a constant which follows from the $g$ time derivative.

In order to solve equation $(\ref{ro-equation})$ we perform the following change of variables
\begin{equation} \label{changvariable}
\bar{z} = \frac{\alpha}{a} e^{-a r};
\rho = F h
\end{equation}
where $F = \left(\frac{\alpha}{a}\right)^{5/2} e^{-\frac{5}{2} a r}$. This give us the well-known Bessel equation of order $\frac{5}{2}$, for the new variable $z$, namely
\begin{equation} \label{bessel1}
\partial^{2} h(\bar{z}) + \frac{1}{\bar{z}} \partial h(\bar{z}) + \left( 1 - \frac{25}{4} \frac{1}{z^{2}}\right) h(\bar{z}) = 0,
\end{equation}
where $\partial = \partial / \partial \bar{z}$. The general solution to this equation is given by
\begin{equation} \label{bessel1sol_z}
h(z) = A J_{\frac{5}{2}} (\bar{z}) + B Y_{\frac{5}{2}} (\bar{z}),
\end{equation}
where $A$ and $B$ are integration constants and $J$, $Y$ are the Bessel function of first and second kind, respectively. In term of the $r$ variable the solution reads
\begin{equation} \label{bessel1sol_r}
\rho(r) = C_{1} e^{-\frac{5}{2} a r} J_{\frac{5}{2}} (\frac{\alpha}{a} e^{-a r}) + C_{2} e^{-\frac{5}{2} a r} Y_{\frac{5}{2}} (\frac{\alpha}{a} e^{-a r}),
\end{equation}
where $C_{1},C_{2}$ are the new constants. In order to obtain the $u$ function \ref{u-variable-sepa1} we need to specify the $g$ function.
In the next section we analyze the case where $g(t)$ is a sine function of time, which  implies in a "standing waves solution" \cite{GOGBERASHVILI1}.

\section{Standing waves solution}

In order to obtain a standing wave solution we choose $u(r,t) = sin(\omega t) \rho(r)$ where $\rho$ is given by Eq. $(\ref{bessel1sol_r})$ and $\omega $ is a constant. In the case $\alpha = \omega$ the solution is
\begin{equation} \label{bessel1sol-sw}
\rho(r) = C_{1} e^{-\frac{5}{2} a r} J_{\frac{5}{2}} (\frac{\omega}{a} e^{-a r}) + C_{2} e^{-\frac{5}{2} a r} Y_{\frac{5}{2}} (\frac{\omega}{a} e^{-a r}).
\end{equation}

Some properties of the function $u(r,t)$ may be found from the solution above. The first one is the fact that the Y function, because its exponential dependence of $r$, do not have divergence on the origin, as one usually find. For large $r$ and $a>0$  the term $e^{-\frac{5}{2} a r} $ in front of $Y$ dominates and we do not have divergence for $r \rightarrow \infty$, as would be expected. For $a<0$, neither $J$ nor $Y$ diverge and we can keep general solution. Furthermore, we require that the phantom-like field is zero on the brane, that is it, at $r = 0$ \cite{GOGBERASHVILI1}. This assumption may be expressed by
\begin{equation}
\label{bound-condic.}
\frac{\omega}{a} =  X_{\frac{5}{2}, n},
\end{equation}
where $X_{\frac{5}{2}, n}$ represents the n-th zero of $J_{\frac{5}{2}}$ or $Y_{\frac{5}{2}}$ depending if we do $C_{1}$ or $C_{2}$ equal to zero in $(\ref{bessel1sol-sw})$. The boundary condition $(\ref{bound-condic.})$ quantizes the $\omega$ frequency.

By this assumption our $u$ function will have zero in some specific $r$ values, denoted by $r_{m}$. For these $r_{m}$ values our model becomes similar to other 6D braneworld models \cite{Gherghetta:2000qi,Giovannini:2001hh,deCarlos:2003nq,Ponton:2000gi,Oda:2000zc,Oda2000a,Koley2007,Gregory:1999gv} as can be seen in the metric $(\ref{metric})$. For $a > 0$ the function \ref{bessel1sol-sw}, for $C_{1} = 0$ or  $C_{2} = 0$, will converge rapidly depending on the value of the ratio $\omega /a$. The quantity of zeros will depends on the value of $a$ and principally on the value of this ratio. In this case we have a finite number of zero. For $a < 0$ (with either $C_{1} = 0$ or $C_{2} = 0)$ the solution will presents an infinity number of zeros.

In order to have a physically accepted scenario, it is necessary that the energy momentum components and the curvature scalar are finite. We have plotted in Fig. (1) the  quantities $\langle T_{xx}(r)\rangle =\langle T_{y y}(r) \rangle = \langle T_{\theta \theta}(r) \rangle$, $ \langle T_{zz}(r) \rangle$, $\langle T_{r r}(r) \rangle$ and $ \langle T_{t t}(r) \rangle$, according to Eq. (\ref{E-M}). In this figure we assume $C_{2} = 0$, $a = A = \kappa = R_{0} = 1$ and $ \omega = 5.76$ which is the first zero for the $J_{5/2}$.  In Fig. (2) we have plotted the same quantities but in this case we have  $\omega = 9.09$, which corresponds to the second zero of $J_{5/2}$. The bulk curvature scalar  $\langle R \rangle$ profile is given in Fig. (3). In this case the frequency $\omega$ assumes the values 5.76, 9.09 and 12.3, which are the first three zeros of $J_{5/2}$. In Fig. (4) we have the 4D curvature scalar $R^{(4)}$ for the same $\omega $ values. The symbol $\langle F \rangle$ represents time average of the function $F$. So the graphics show the time average profiles of the energy momentum tensor components $T_{MN}$ and the curvature scalars of the bulk and brane, $R$ and $R^{(4)}$, respectively.  In this context, as will be seen in the next section, the quantity $\langle e^{b u} \rangle$ is given by
\begin{equation} \label{exp_average}
\langle e^{b u} \rangle = I_{0} (b \rho(r)),
\end{equation}
with $\rho (r)$ given by Eq. (\ref{bessel1sol-sw}), where $b$ is any constant and $I_{0}$ is the modified Bessel function of order zero.

The figures show that the model proposed in this work produces a scenario where none of this important quantities are infinite. In Fig. $(1)$ and Fig. $(2)$ we have a sketch of the time average components of the energy momentum tensor. In the first case, one has $\omega = 5.76$ and in Fig. (2), $\omega = 9.09$. The solid line represents $ \langle T_{t t}(r) \rangle$, the dotted line represents $ \langle T_{zz}(r) \rangle$, the dashed line represents $\langle T_{r r}(r) \rangle$ and finally the dash-dotted line represents $\langle T_{xx}(r)\rangle =\langle T_{y y}(r) \rangle = \langle T_{\theta \theta}(r) \rangle$. As can be seen from these figures, the energy density, depending on the value of the ratio $\omega /a$,  oscillates between positive and negative values, but the pressure is ever negative. The positive energy density with negative pressure is a feature of the phantom energy \cite{Caldwell}, but in this case the energy density assumes negative values which shows the exotic
nature of the source which is not a phantom field in the sense that is studied in the literature \cite{Caldwell,Merab2011a,Merab2011b,Merab2012}. The dominant energy condition is not satisfied, which is common in phantom field theory, because, even in the case that the energy density is positive, it would be necessary it to be greater than the pressure components. However, all of these quantities are finite which is less problematic in comparison with infinity unbounded energy momentum components. It is applicable to say that the presence of models which do not satisfy the dominant energy condition is frequent in the literature. We may cite the 5D version  of this present model \cite{GOGBERASHVILI1} as a case where the energy conditions is not satisfied. In 6D braneworld there exist models that present this unorthodox feature, as the proposed by Koley and Kar \cite{Koley2007} and the thin string-like brane \cite{Gherghetta:2000qi,Tinyakov:2001jt}.

In Fig. $(3)$ we plot the bulk curvature scalar for three different values of $\omega$: dashed line for $\omega = 5.76$; dotted line for $\omega = 9.09$ and solid line for $\omega = 12.3$. We see that in all these cases $\langle R \rangle$ is finite and positive, which reveals a dS scenario. Beside the fact that the bulk is a de Sitter spacetime, the brane is an AdS universe as can be seen in Fig. (4). In this figure the values assumed by $\omega$ are: dashed line for $\omega = 5.76$; dotted line for $\omega = 9.09$ and solid line for $\omega = 12.3$.

Therefore, we can conclude from all these figures that the brane behaves like a thick brane which is located in the origin and extends itself to the position $r = 1.5$, approximately. This profile and the behaviour of the energy components are consequences of the exotic source that generate it. On the other hand, it is worthwhile to stress that these figures represent time-average of the quantities in question. Hence, it is possible, in some time interval, have the energy conditions satisfied.

\begin{figure}[htb] 
       \begin{minipage}[b]{0.48 \linewidth}
           \fbox{\includegraphics[width=\linewidth]{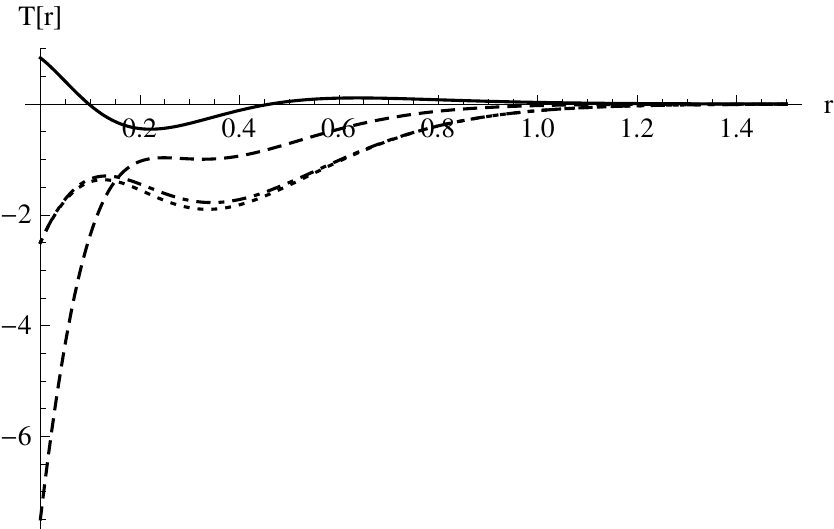}}\\
           \caption{\it $\langle T \rangle$ profile, $\omega = 5.76$, $ \langle T_{t t} \rangle$ (solid line), $ \langle T_{zz} \rangle$ (dotted line), $\langle T_{r r} \rangle$ (dashed line), $ \langle T_{\theta \theta} \rangle$ (dash-dotted line).}
           \label{fig:1}
       \end{minipage}\hfill
       \begin{minipage}[b]{0.48 \linewidth}
           \fbox{\includegraphics[width=\linewidth]{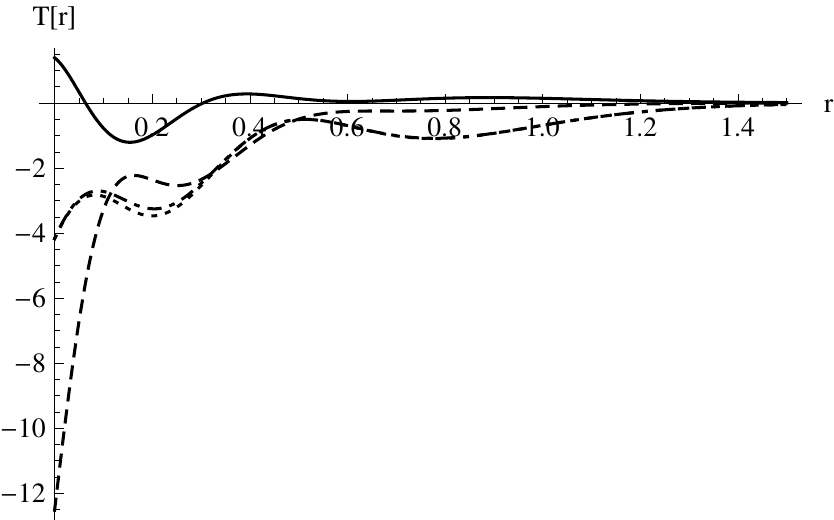}}\\
           \caption{\it $\langle T \rangle$ profile, $\omega = 9.09$, $ \langle T_{t t} \rangle$ (solid line), $ \langle T_{zz} \rangle$ (dotted line), $\langle T_{r r} \rangle$ (dashed line), $ \langle T_{\theta \theta} \rangle$ (dash-dotted line).}
           \label{fig:2}
       \end{minipage}
   \end{figure}

\begin{figure}[htb] 
       \begin{minipage}[b]{0.48 \linewidth}
           \fbox{\includegraphics[width=\linewidth]{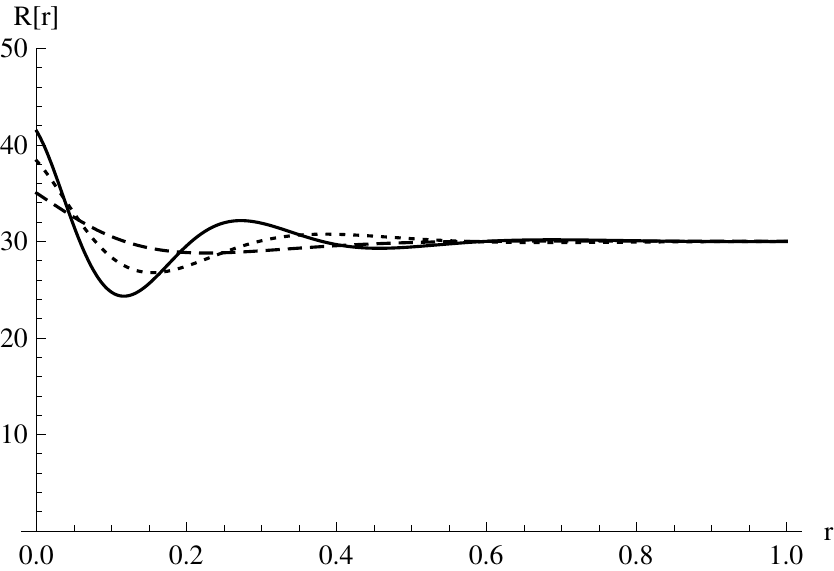}}\\
           \caption{\it $\langle R \rangle$ profile. $\omega$ = 5.76 (dashed line), 9.09 (dotted line), 12.3 (solid line).}
           \label{fig:3}
       \end{minipage}\hfill
       \begin{minipage}[b]{0.48 \linewidth}
           \fbox{\includegraphics[width=\linewidth]{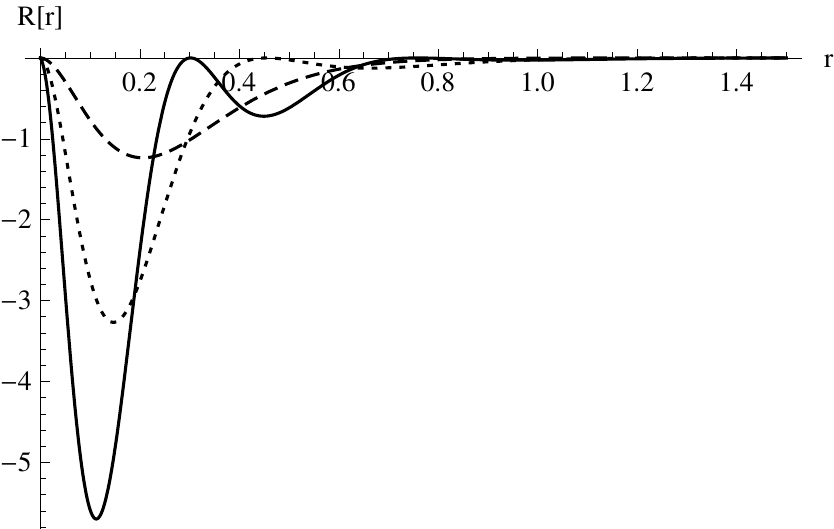}}\\
           \caption{\it $\langle R^{(4)} \rangle$ profile. $\omega$ = 5.76 (dashed line), 9.09 (dotted line), 12.3 (solid line).}
           \label{fig:4}
       \end{minipage}
   \end{figure}

In braneworld scenarios there is an interest in the potential of the model to localize the Standard Model (SM) fields. However, to find a model where all SM fields are localized only by means of gravitational interaction (without the necessity of additional fields, as the dilaton field) is not an easy task. The simple extension of the RS model from 5D to 6D sometimes is sufficient to eliminate the necessity of additional interaction in some cases. Indeed, as can be seen in the works of references \cite{Tahim2009,Kehagias2001}, we have 5D models where the localization of the gauge and Kalb-Ramond fields, respectively, are possible only with the introduction of the dilaton field  but in 6D the same fields are localized only by means of gravity interaction \cite{Oda:2000zc, Oda2000a,Luis2012}. Here, we introduce a six dimensional model which generalizes both a 6D thin braneworld model previously studied \cite{Gherghetta:2000qi,Giovannini:2001hh,Ponton:2000gi,Tinyakov:2001jt,Oda:2000zc, Oda2000a,Luis2012,Gregory:1999gv} and also a so called 5D "standing waves" braneworld model \cite{GOGBERASHVILI1,Merab2011a,Merab2011b}, where it is possible to localize gravity, scalar, vector and tensor fields but it was not possible to trap right fermions \cite{Merab2012}. We argue that our synthesis model enables the SM fields trapping without the necessity of any other interaction than gravity. In this work, we shall give the first step in this analysis beginning by the scalar field whose localization will be studied in the next section.

\section{Scalar field localization}

In order to study the localization of the bulk scalar field consider the action
\begin{equation} \label{scalar-action}
S = -\frac{1}{2} \int{d^{6}x \sqrt{-g}g^{MN}\partial _{M} \Phi \partial _{N} \Phi},
\end{equation}
from which we derive the equation of motion
\begin{equation}
\label{scalar_eq_mot}
\frac{1}{\sqrt{-g}}\partial _{M}\left( \sqrt{-g}g^{MN}\partial_{N}\Phi\right) = 0.
\end{equation}
From the metric \textit{ansatz} $(\ref{metric})$ we have that $\sqrt{-g} = e^{5 a r}$. Thus, Eq. $(\ref{scalar_eq_mot})$ may be rewritten  as
\begin{equation}
\label{scalar_eq_mot1}
\left[ \partial _{t} ^{2} - e^{-u}\left( \partial _{x} ^{2} + \partial _{y} ^{2}\right) - e^{3u}\partial _{z} ^{2} - \frac{e^{-u}}{R_{0} ^{2}}\partial _{\theta} ^{2}\right] \Phi = e^{-3a r} \left( e^{5 a r} \Phi ^{'}\right)^{'}.
\end{equation}

Consider a solution of the form
\begin{equation}
\label{scalar_var_sep}
\Phi (x^{M})= \Psi(r, t) \chi(x, y) \zeta(z) e^{il\theta}.
\end{equation}

Substituting Eq. $(\ref{scalar_var_sep})$ in Eq. $(\ref{scalar_eq_mot1})$, we obtain the following system of equations
\begin{equation}
\label{syst1}
\left( \partial _{x} ^{2} + \partial _{y} ^{2}\right) \chi + \left( p_{x} ^{2} + p _{y} ^{2}\right) \chi = 0,
\end{equation}
\begin{equation}
\label{syst2}
\partial _{z} ^{2} \chi + p_{z} ^{2}\chi = 0,
\end{equation}
\begin{equation}
 \label{syst3}
\left[ \partial _{t} ^{2} + e^{-u}\left( p _{x} ^{2} + p _{y} ^{2}\right) + e^{3u}p _{z} ^{2} + \frac{l^{2}e^{-u}}{R_{0} ^{2}}\right] \Psi = e^{-3a r} \left( e^{5 a r} \Psi ^{'}\right)^{'}.
\end{equation}

Now we turn back to Eq. $(\ref{syst3})$ and we separate the variables by making $\Psi (r,t) = e^{iEt}\bar{\rho}(r)$,  which will give us
\begin{equation}
\label{syst4}
\left( e^{5ar} \bar{\rho}(r) ^{'}\right)^{'} - e^{3ar} G(r) \bar{\rho}(r) = 0,
\end{equation}
where
\begin{equation}
\label{Gr}
G(r) = \left( p _{x} ^{2} + p _{y} ^{2}\right)\left(e^{-u} -1\right) + p _{z} ^{2} \left(e^{3u} -1\right) + \frac{l^{2}}{R_{0} ^{2}} e^{-u}.
\end{equation}

For further convenience we write $(\ref{syst4})$ in an analogue non-relativistic quantum mechanical problem by making the variable change $\bar{\rho}(r) = e^{-5/2 a r}\bar{\Psi} (r)$. Therefore, we obtain
\begin{equation}
\label{schr-like}
\bar{\Psi}^{''}(r)  - V(r) \bar{\Psi}(r) = 0,
\end{equation}
where
\begin{equation} \label{Pot-schr}
V(r) = \frac{25}{4} a^{2} + e^{-2ar} G(r).
\end{equation}

In order to analyze the localization of the scalar field we need to obtain the rt-dependent function $\Psi$ in $(\ref{syst3})$. We shall do it by means of equation Eq. (\ref{schr-like}), but this will be done only for the scalar zero mode and s-wave. So we will assume $(l=0)$ and $E = p _{x} ^{2} + p _{y} ^{2} + p _{z} ^{2}$. Furthermore, we shall consider the case where $\omega >> E$ which allows us to perform the time-averaging of $V(r)$ reducing the problem to just one variable, $r$. Then, using the following expressions
\begin{equation}
\label{exp-serie}
e^{bu} = \sum_{n = 0} ^{+ \infty} \frac{(bu)^{n}}{n!},
\end{equation}
\begin{equation}
\label{time-aver}
\omega /2 \pi \int_{0} ^{2 \pi / \omega} [sin (\omega t)]^{m} = 2^{-2n} (2n)! (n!)^{-2}; (m = 2n),
\end{equation}
we find
\begin{equation}
 \label{exp-serie-solution}
\left\langle e^{bu} \right\rangle = 1 + \sum_{n = 1} ^{+ \infty} \frac{(b)^{2n}}{2^{2n}(n!)^{2}} [ C_{1} e^{-\frac{5}{2} a r} J_{\frac{5}{2}} (\frac{\omega}{a} e^{-a r}) + C_{2} e^{-\frac{5}{2} a r} Y_{\frac{5}{2}} (\frac{\omega}{a} e^{-a r})
]^{2n} = I_{0} (b \rho(r)),
\end{equation}
where $I_{0}$ is the modified Bessel function of the first kind. Even in this case, as can be seen from the expression $(\ref{exp-serie-solution})$, to find analytical solution for Eq. $(\ref{schr-like})$ is not so easy. We shall accomplish it considering asymptotic approximations.

Note that for $C_{2} = 0$ in Eq. $(\ref{bessel1sol-sw})$, the $u(r,t)$ will depends on first kind Bessel function $J_{\sqrt{\frac{5}{2}}}$. The expansion \ref{exp-serie-solution} now will be given by
\begin{equation} \label{exp-serie-solution-J}
\left\langle e^{bu} \right\rangle = 1 + \sum_{n = 1} ^{+ \infty} \frac{(bC_{1})^{2n} e^{-5 a n r}}{2^{2n}(n!)^{2}} [J_{\frac{5}{2}} (\frac{\omega}{a} e^{-a r})]^{2n}.
\end{equation}

Let us study the behavior of equation (\ref{schr-like}) in two distinct regions: far from and near to the brane.
For $r \rightarrow + \infty$ the arguments in $J_{\frac{5}{2}}$ goes to zero ($(\omega /a)e^{-a r} \rightarrow 0$), so the expression $(\ref{exp-serie-solution-J})$ will be approximated as $\left\langle e^{bu} \right\rangle \approx 1$. Consequently equation $(\ref{schr-like})$ will assume the simple form
\begin{equation}
 \label{schr-like-aproxJ}
\bar{\Psi}^{''}(r)  - \frac{25}{4} a^{2} \bar{\Psi}(r) = 0,
\end{equation}
whose solution is $e^{\pm \frac{5}{2} a r}$. We choose $\bar{\Psi} = e^{- \frac{5}{2} a r}$ and $a > 0$ which is naturally convergent. This solution is the same find in 5D standing-wave, for the localization of scalar field, in the same asymptotic limit considered here \cite{Merab2011b}.

For $r \rightarrow 0$ the equation (\ref{schr-like}) may be approximated by
\begin{equation}
 \label{schr-like-small-r}
\bar{\Psi}^{''}(r)  -\left( \frac{65}{2} c a^{2} r^{2} - 12 c a r + c^{'}\right) \bar{\Psi}(r) = 0.
\end{equation}
This is a generalization of the same equation found for localization of scalar field in 5D standing waves braneworld \cite{Merab2011b}. The constants $c$ and $c^{'}$ are given, respectively, by
\begin{equation}
 \label{constant1}
c = \left(\frac{C_{1} }{\Gamma (\frac{7}{2})}\right)^{2} \left(\frac{\omega }{2 a}\right)^{5} (p_{x} ^{2} + p_{y} ^{2} + 9 p_{z} ^{2}),
\end{equation}
and
\begin{equation}
\label{constant2}
c^{'} = \frac{25}{4} a^{2} + c.
\end{equation}

Note that equation $(\ref{schr-like-small-r})$ is nothing but the parabolic cylinder equation whose general solution is
\begin{eqnarray}
\label{sol_schr-like-small-r}
\bar{\Psi} (r)  & = &  E_{1} D_{\mu} \left( -\frac{12 (2 c)^{1/4}}{\sqrt{a \sqrt{65^{3}}}} + \left(\sqrt{a \sqrt{130 c}}\right) r \right)\nonumber\\
                & + &  E_{2} D_{\nu} \left( -i \frac{12 (2 c)^{1/4}}{\sqrt{a \sqrt{65^{3}}}} + i \left(\sqrt{a \sqrt{130 c}}\right) r \right).
\end{eqnarray}
where $D$ is the parabolic cylinder function and $E_{1}, E_{2}$ are integration constants. We see that $E_{2}$ must be zero in order to have a real solution. The $\mu$, $\nu $ indexes are given respectively by
\begin{equation}
\label{D-index1}
\mu = -\frac{64 a \sqrt{130 c} - 144 c + 130 c^{'}}{130 a \sqrt{130 c}},
\end{equation}
and
\begin{equation}
\label{D-index2}
\nu = \frac{- 4225 a \sqrt{ c} - 72 \sqrt{130} c + 65 \sqrt{130} c^{'}}{8450 a \sqrt{ c}}.
\end{equation}

For $E_{2} = 0$ and $\omega/a = 5,76$ it is possible to choose the overall constants in order to expand our solution (\ref{schr-like-small-r}) as
\begin{equation}
\label{schr-like-serie}
\bar{\Psi} (r) \simeq 2.83 - 18.7 r + 5.58 r^{2} - 9.78 r^{3} + 10.7 r^{4} - 0[r]^{5}.
\end{equation}
In this case the function   $(\ref{schr-like-small-r})$ assumes the form $D_{-0.53}(-1.5 + 2.85 r)$. The approximation   $(\ref{schr-like-serie})$ and, consequently, the extra part of the scalar zero-mode wave function $\bar{\rho}(r)$ has a maximum at $r = 0$ and it falls off from the brane, assuming the asymptotic form  $e^{- 5 a r}$ which is in accordance with Ref. \cite{Merab2011b}.

This result shows the localization of the scalar field zero-mode in this model. It is interesting to point out that the localization is obtained for an increasing warp factor whereas in the thin string-like brane the localization of the zero-mode scalar field is obtained for a decreasing warp factor \cite{Oda:2000zc,Oda2000a}.

\section{Remarks and conclusions}

We have constructed a six-dimensional standing-wave model where the brane is generated from a phantom-like scalar field. We have supposed the compact dimension staying on the brane and small enough to accomplish an realistic model. The metric ansatz, unlike the most braneworld models present in the literature, is anisotropic and non-static and the compact dimension is time dependent. The bulk geometry tends asymptotically to spacetime with positive cosmological constant. From the action composed by the gravity and a phantom-like field we derived the equation of motion and we have evaluated the phantom scalar expression.

The exotic source which generates the brane is well characterized by its energy density. As can be seen in Fig. (1) and Fig. (2), this quantity has not a fixed value varying in intensity and signal. While the "traditional" phantom field has negative energy density, in the $6$D standing-wave model this quantity may be positive and it is finite. This property prevents the model from unbounded infinite energy density. Far from the brane the bulk  presents the features expected for a space with non-null cosmological constant, as expected. Once we know the importance of phantom fields in cosmology, but in the same time, the difficulty  that the phantom physics suffers, like the negative unbounded energy density, our model may appears as an alternative phantom model for phenomenology.

Moreover, in the context of braneworld it is suitable to investigate whether a model is able to
localize fields. In order to analyze if our solution allows field
localizations we studied the zero mode scalar field localization. The results have shown that there is a
zero mode localization for the scalar field. The solution found here is in accordance with the one first encountered in five dimensions \cite{Merab2011b}.

In future works we intend to investigate the second part of the general solution to our phantom-like scalar field. Further, we shall verify the possibility of a brane with a decreasing warp factor. Moreover, we shall study the localization of the overall SM fields, particularly the right-handed fermions which are not localized in the 5D version of this model,as was previously mentioned.

\section{Acknowledgments}

The authors thanks the financial supporting of CNPq and CAPES (Brazilian Agencies).


\end{document}